\documentclass{article}
\usepackage{graphicx}
\usepackage{amsmath}

\begin{document}

\title{A non-local hidden-variable model that violates Leggett-type inequalities}
\author{F. De Zela \\
%EndAName
Departamento de Ciencias, Secci\'{o}n F\'{i}sica \\
Pontificia Universidad Cat\'{o}lica del Per\'{u}, Apartado 1761, Lima,
Per\'{u}.}
\maketitle

\begin{abstract}
Recent experiments of Gr\"{o}blacher \emph{et al.} proved the violation of a
Leggett-type inequality that was claimed to be valid for a broad class of
non-local hidden-variable theories. The impossibility of constructing a
non-local and realistic theory, unless it entails highly counterintuitive
features, seems thus to have been experimentally proved. This would bring us
close to a definite refutation of realism. Indeed, realism was proved to be
also incompatible with locality, according to a series of experiments
testing Bell inequalities. The present paper addresses the said experiments
of Gr\"{o}blacher \emph{et al.} and presents an explicit, contextual and
realistic, model that reproduces the predictions of quantum mechanics. It
thus violates the Leggett-type inequality that was established with the aim
of ruling out a supposedly broad class of non-local models. We can thus
conclude that plausible contextual, realistic, models are still tenable.
This restates the possibility of a future completion of quantum mechanics by
a realistic and contextual theory which is not in a class containing only
highly counterintuitive models. The class that was ruled out by the
experiments of Gr\"{o}blacher \emph{et al.} is thus proved to be a limited
one, arbitrarily separating models that physically belong in the same class.

\end{abstract}

\section{Introduction}

Recently, Gr\"{o}blacher \emph{et al} \cite{gröblacher} reported experiments
performed with two entangled photons, showing the violation of a
Leggett-type inequality that was derived by the said authors. According to
them, such a violation rules out a broad class of hidden-variable (HV)
models based on non-local realism. This would be an important step towards
answering the question about the completeness of quantum mechanics (QM), a
question that was raised by Einstein, Podolsky and Rosen in their celebrated
paper of 1935. Several experiments testing Bell inequalities served the
purpose of closing the detection and the locality loopholes, and for many -
though not for all \cite{santos2} \ - researchers in the field, ``it is
reasonable to consider the violation of local realism a well established
fact'' \cite{gröblacher}. Accordingly, a completion of QM through a HV
theory would require that this theory be based upon the concept of
non-locality. Theories of this kind are sometimes taken on an equal basis as
``contextual'' theories and, in fact, the Leggett-type inequality that was
tested by Gr\"{o}blacher \emph{et al} does not distinguish contextuality
from non-locality. It simply admits as a possibility that measurement
outcomes may depend on parameters in distant regions, without specifying
whether the involved separations are space-like or time-like. The class
defined by the Leggett-type inequality may include certain non-local
realistic theories; but the experiments performed by Gr\"{o}blacher \emph{et
al} were not able to distinguish these theories from local contextual ones.
Indeed, for the events involved (photon detection or emission) the time
coordinate is not registered and the settings remain fixed. Hence, one
cannot tell whether space-time intervals $\Delta s^{2}\equiv c^{2}\Delta
t^{2}-\Delta x^{2}-\Delta y^{2}-\Delta z^{2}$ between pairs of events are
time-like, space-like, or light-like, excepting the restricted class of
simultaneous events ($\Delta t=0$, e.g., coincidences in photon detections),
for which they are space-like. Strictly speaking, there is a difference
between ``non-locality'' and ``contextuality''. ``Non-locality'' is often
understood in a relativistic sense (causal influences can travel faster than
light), while ``contextuality'' refers to the fact that a measurement
outcome or the state of a system may generally depend on the context in
which the measurement takes place or the system is prepared \footnote{%
The term ``non-local'' is nevertheless often used in a non-relativistic
sense also, for example when referring to quantum correlations, these
correlations being not necessarily attributed to superluminal causal
influences.}. Now, the ambiguity concerning ``non-local'' and ``contextual''
models can be removed if we ask the model-maker to consider changes in the
context and to tell after how much time these changes should manifest
themselves in the measurement results. Whatever the answer, the model will
be either local or non-local. Note that ``contextual-local'' theories are
not excluded from the outset, as it might appear at first sight. Indeed, a
relativistic theory may admit contextuality if it requires fixing boundary
conditions to solve its fundamental equations. Physically, one assumes that
these conditions have been fixed well in advance so as to allow causal
influences to propagate, without conflicting relativity, from the boundaries
towards those parts of the system that are the subject of the fundamental
equations. If we instead consider boundary conditions that change with time,
then we are faced with the choice between locality and non-locality. Here we
will mostly use the term contextuality rather than non-locality, because the
cases we shall address do not distinguish whether the space-time intervals
between events (e.g., measurements) are time-like or space-like. The
recently reported experiments \cite{gröblacher} are said to exclude a broad
class of non-local HV models. Because the experimental arrangement of
Gr\"{o}blacher \emph{et al.} could not distinguish local from non-local
influences, we prefer to use the term ``contextual'' rather than
``non-local'' when referring to it. Anyhow, Gr\"{o}blacher et al. arrived at
the conclusion that any possible completion of QM through a realistic HV
theory would be a highly counterintuitive one, entailing exotic features
such as a departure from Aristotelian logic, actions into the past, etc.
Here we show that the class of contextual models addressed in the
experiments of Gr\"{o}blacher \emph{et al.} is - from a physical point of
view - a rather limited one. Indeed, as we shall see, this class arbitrarily
excludes a type of contextual models that should have been addressed, for
the approach to be physically self-consistent. The exclusion of these models
invalidates, in fact, the claim of Gr\"{o}blacher \emph{et al.} To show
this, we will present an explicit, contextual model that violates the
Leggett-type inequality that the reported experiments put to the test. Of
course, the Leggett-type inequality remains true for the restricted class
considered by Gr\"{o}blacher \emph{et al}. Our claim is that this class is
physically unsound, for the following reason. Gr\"{o}blacher \emph{et al.}
derived a Leggett-type inequality for a class of non-local models in which
there are two measuring devices (e.g., polarizers) and a source (of
entangled pairs). Furthermore, following Leggett, they relaxed the locality
assumption for the polarizers, so that measurement outcomes at one device
could depend on the settings of the other. However, the source was supposed
to be somehow ``protected'' from any non-local influences, a condition that
appears to be physically unsound once we have relaxed the locality
assumption. In the present case, we will assume non-locality for the source
as well as for the polarizers. To be sure, Leggett's original assumption of
a source being exclusively influenced by conditions on its neighborhood
could be - in principle - experimentally realized, as we will discuss next,
but the experiments of Gr\"{o}blacher \emph{et al.} were not designed to
fulfill this requirement.

Leggett's assumption of a ``protected'' source, referred to above, could be
fulfilled in experiments with variable polarizers. Such experiments could
test non-local models besides contextual ones. Let us first consider testing
contextual local hidden-variable models. Within such a framework we could
conceive an arrangement for which the settings of the polarizers constitute
events that are space-like separated from the emission at the source. In
this way the source becomes effectively ``protected'' from non-local
influences and the distribution describing it (see below) can be assumed to
be independent of a context that is beyond its immediate neighborhood. Let
us consider next non-local models. In this case, though we abandon the
framework of special relativity, it is still reasonable to assume that we
may define an absolute future with respect to a given event in some
reference frame, e.g., the laboratory. Under this assumption, we could again
``protect'' the source from non-local influences, by making sure that the
settings of the polarizers be in the absolute future of the emission at the
source. Admittedly, it would be extremely difficult to perform experiments
of this kind, particularly in those cases in which the emission at the
source constitutes an stochastic process. Nevertheless, as a matter of
principle, it is not quite unreasonable to assume that the emission at the
source be uninfluenced by the settings of distant polarizers.

Finally, we should stress that the ``free will assumption'' is not an issue
here, for similar reasons as the ones just discussed. According to the free
will assumption, an operator is free to choose between different measurement
settings \cite{kofler}, i.e., choices are not predetermined by some initial
conditions. We will consider experimental situations like the one addressed
by Gr\"{o}blacher \emph{et al}, in which the settings of the instruments are
not changed during the flight of the particles. If these settings are fixed
``sufficiently in advance to allow them to reach some mutual rapport by
exchange of signals with velocity less than or equal to that of light'' \cite
{bell}, contextual models - as already said - may well be in accordance with
relativity. If, on the contrary, the settings were changed by an operator
during the flight of the particles, then contextual models could violate
relativistic causality. A way to avoid this possible conflict with
relativity is to relax the free will assumption and consider that the actual
settings of the instruments by an operator have been fixed in the past by
some initial conditions, thereby making free will a mere illusion. The
models considered here do not address the question of free will, because
they assume fixed settings.

The paper is organized as follows. Section \ref{section1} begins with a
discussion of the experiment of Gr\"{o}blacher \emph{et al}. In subsection
\ref{subsection1} we introduce the Kochen-Specker model for a single qubit
in a form which is amenable to generalization. Building upon this model we
present in subsection \ref{subsection2} an extension that works for two
qubits. This is the central result of the paper. Finally, in section \ref
{section2} we present our conclusions.

\section{The Kochen-Specker model and its extension to the nonlocal case%
\label{section1}}

Gr\"{o}blacher \emph{et al} \cite{gröblacher} carried out experiments with
two entangled qubits which were realized as polarization entangled photons
generated via spontaneous parametric down-conversion. The contextual models
that were put to the test should satisfy the following assumptions: (1)
measurement-outcomes reveal pre-existing properties (realism \footnote{%
I use the term ``realism'' as it is defined by Gr\"{o}blacher \emph{et al}.
Alternatively, for some authors ``realism'' can be defined so as to allow
that measurement results be generated at the moment of measurement.}); (2)
physical states are mixtures of subensembles with definite properties (e.g.,
polarization); (3) the expectation values taken for each subensemble
coincide with the predictions of QM (e.g., polarization-states obey Malus'
law).

Consider a source emitting pairs of photons towards two measuring devices
whose respective settings, as given by unit vectors $n_{a}$ and $n_{b}$, are
fixed by Alice and Bob in each run of the experiment. The emitted photons
have well-defined polarizations $n_{u}$ and $n_{v}$. Let us denote by $A$
and $B$ the outcomes for polarization measurements along $n_{a}$ and $n_{b}$%
, respectively. Their values are $\pm 1$, corresponding to
transmission/absorption of a photon. By writing, e.g., $A=A(\lambda
,n_{a},n_{b})\equiv A(\lambda ,a,b)$ we make explicit reference to the
assumption that Alice's results may depend on Bob's settings and that the
model is, thus, contextual. Let $\rho _{uv}(\lambda )$ be the subensemble
distribution of the photon pairs emitted by the source with polarizations $%
n_{u}$ and $n_{v}$. It is required that, according to Malus' law, local
averages satisfy

\begin{eqnarray}
\overline{A} &=&\int d\lambda \rho _{uv}(\lambda )A(\lambda
,a,b)=n_{u}.n_{a},  \label{1} \\
\overline{B} &=&\int d\lambda \rho _{uv}(\lambda )B(\lambda
,b,a)=n_{v}.n_{b}.  \notag
\end{eqnarray}
For a source emitting well-polarized photon pairs the correlation function
of measurement results is given by $\overline{AB}(u,v)=\int d\lambda \rho
_{u,v}(\lambda )A(\lambda ,a,b)B(\lambda ,b,a)$. For a more general source
we assume to have at our disposal a distribution function $F(u,v)$ that
describes the mixture of polarized pairs produced by the source. The general
correlation function is thus given by $E_{ab}=\left\langle AB\right\rangle
=\int dudvF(u,v)\overline{AB}(u,v)$. So far, we have followed the
assumptions and, closely, the notation of Gr\"{o}blacher \emph{et al.} These
authors derived a Leggett-type inequality to be experimentally tested. The
said inequality follows from the identities $-1+\left| A+B\right|
=AB=1-\left| A-B\right| $, which are fulfilled whenever $A=\pm 1$ and $B=\pm
1$. Multiplying these identities by $\rho _{uv}(\lambda )$, integrating over
$\lambda $ and noting that $\overline{\left| A\pm B\right| }\geq \left|
\overline{A}\pm \overline{B}\right| $, one obtains $-1+\left| \overline{A}+%
\overline{B}\right| \leq \overline{AB}\leq 1-\left| \overline{A}-\overline{B}%
\right| $. Further multiplication by $F(u,v)$ and subsequent integration
over $(u,v)$ leads to a Leggett-type inequality that puts a bound on the
correlations $E_{ab}$ predicted by a contextual HV model. QM predicts $%
E_{ab}=-n_{a}\cdot n_{b}$ and these values violate the said inequality. The
experiments of Gr\"{o}blacher \emph{et al.} reproduce the quantum-mechanical
predictions with high accuracy. Now, as we observe from the above
definitions, contextuality has been restricted to the measurement-outcomes
of Alice and Bob, as if the source could not be affected by, e.g., Bob's
settings, while $A$ could be affected by such settings. This is, of course,
an arbitrary assumption from the viewpoint of a contextual model. All the
more, if we remind ourselves that what is considered a ``source'' (or part
of it) in one experiment, may be dubbed as ``measuring device'' in the
other. A contextual model should consistently assume functions of the form $%
\rho _{uv}(\lambda ,a,b)$ and $F(u,v,a,b)$. Though the derivation of the
Legget-type inequality presented in Ref.\cite{gröblacher} appears to be
insensitive to a possible contextuality of $\rho _{uv}$, a distribution
function of the form $F(u,v,a,b)$ invalidates the derivation of the
Legget-type inequality of Ref.\cite{gröblacher} (see Supplementary
Information to Ref.\cite{gröblacher}, and \cite{leggett}). Hence, the
contextual models that have been tested in the experiments of Gr\"{o}blacher
\emph{et al.} belong to a restricted class. This class would be defined
through the nonphysical assumption that sources and measuring devices are
differently influenced by the experimental context, the sources being
somehow ``protected'' from contextual influences. In fact, it is possible to
construct an explicit contextual model that reproduces the experimental
results reported in Ref.\cite{gröblacher}, as we shall see next. The
construction is based on the Kochen-Specker (KS) model for a single qubit
\cite{kochen,rudolph}. We will first discuss the KS\ model, but in a way
that departs from its original formulation. Our approach has been tailored
in a way that can be generalized to the two-qubit case.

\subsection{The Kochen-Specker model for a single qubit\label{subsection1}}

In the KS model, the HVs $\lambda $ span the unit sphere $S^{2}$. Thus, $%
\lambda $ can be parameterized as a unit vector $n_{\lambda }=(\sin \theta
_{\lambda }\cos \varphi _{\lambda },\sin \theta _{\lambda }\sin \varphi
_{\lambda },\cos \theta _{\lambda })$. Henceforth we will write,
interchangeably, $\lambda $ or $n_{\lambda }$, for these and all other unit
vectors playing the role of HVs. Now, $S^{2}$ is also the Bloch sphere,
which serves to represent geometrically a qubit $\left| \psi \right\rangle
=\cos \left( \theta _{\psi }/2\right) $ $e^{-i\varphi _{\psi }/2}\left|
+\right\rangle +\sin \left( \theta _{\psi }/2\right) $ $e^{i\varphi _{\psi
}/2}\left| -\right\rangle $, where $\left| \pm \right\rangle $ are the $%
S_{z} $-eigenvectors. Indeed, we can assign to each $\left| \psi
\right\rangle $ a unit vector $n_{\psi }=(\sin \theta _{\psi }\cos \varphi
_{\psi },\sin \theta _{\psi }\sin \varphi _{\psi },\cos \theta _{\psi })$ on
the Bloch sphere and, reciprocally, to each unit vector $n_{\psi }\in S^{2}$
it corresponds a normalized qubit-state $\left| \psi \right\rangle $ (modulo
a phase). The KS model assigns to each qubit $\left| \psi \right\rangle $ a
probability density $\rho _{\psi }(\lambda )d\lambda $, with $d\lambda $
being a suitable measure. This corresponds to a completion through HVs of
the supposedly ``incomplete'' description of physical reality that is
provided by the state-vector $\left| \psi \right\rangle $. Consider now
Alice's observable $\widehat{A}$. As is well known, it can be written in the
form $\widehat{A}=a_{0}+\mathbf{a}\cdot \mathbf{\sigma }$, where $\mathbf{%
\sigma }$ stands for the triple of Pauli matrices. The eigenvectors of $%
\widehat{A}$ will be denoted by $\left| \psi _{a}^{\pm }\right\rangle $.
They are the same as those of $n_{a}\cdot \sigma $, with $n_{a}=\mathbf{a}/|%
\mathbf{a}|$. To projection operators like $\widehat{\Pi }_{a}^{\pm }=\left|
\psi _{a}^{\pm }\right\rangle \left\langle \psi _{a}^{\pm }\right| $ we
assign characteristic dichotomic functions, $\chi _{a}^{\pm }(\lambda )$,
each taking the values $1$ and $0$, in correspondence to whether a given
event does or does not take place. For example, $\chi _{a}^{+}(\lambda )$
states whether a measurement of $\widehat{A}$ along $n_{a}$ does or does not
produce a ``positive'' result (e.g., detection along the upward direction in
a Stern-Gerlach set-up), and similarly for $\chi _{a}^{-}(\lambda )$
(counting now as ``positive'' a detection along the downward direction in
the Stern-Gerlach set-up). That is, if $\chi _{a}^{+}(\lambda )=1$, then $%
\chi _{a}^{-}(\lambda )=0$, and viceversa. Hence, we can write $A(\lambda
)=\chi _{a}^{+}(\lambda )-\chi _{a}^{-}(\lambda )$. The functions $\rho
_{\psi }(\lambda )d\lambda $ and $\chi _{a}^{\pm }(\lambda )$ should be
chosen so as to afford, for all $\left| \psi \right\rangle $, that

\begin{equation}
\left\langle \psi \right| \widehat{\Pi }_{a}^{\pm }\left| \psi \right\rangle
=\int \rho _{\psi }(\lambda )\chi _{a}^{\pm }(\lambda )d\lambda .  \label{1a}
\end{equation}
The model then reproduces all quantum-mechanical predictions about
probabilities of measurement outcomes, in the sense that for all qubits $%
\left| \psi \right\rangle $ and for any operator $\widehat{A}$, it holds
true that $\left\langle \psi \right| \widehat{A}\left| \psi \right\rangle
=\int \rho _{\psi }(\lambda )A(\lambda )d\lambda $. This suffices for our
present purposes, although, if required, we could also prescribe the
dynamics between measurements. Eq.(\ref{1a}) will be satisfied if for any $%
\left| \psi _{a}\right\rangle $ and $\left| \psi _{b}\right\rangle $, such
that $n_{i}\cdot \sigma \left| \psi _{i}\right\rangle =\left| \psi
_{i}\right\rangle $, $i=a,b$, we define $\rho _{a}(\lambda )d\lambda $ and $%
\chi _{b}(\lambda )$ so that $\left| \left\langle \psi _{b}\mid \psi
_{a}\right\rangle \right| ^{2}=\left\langle \psi _{a}\right| \widehat{\Pi }%
_{b}\left| \psi _{a}\right\rangle =\left\langle \psi _{b}\right| \widehat{%
\Pi }_{a}\left| \psi _{b}\right\rangle =\cos ^{2}\left( \theta
_{ab}/2\right) =\int \rho _{a}(\lambda )\chi _{b}(\lambda )d\lambda $, where
$\theta _{ab}=\cos ^{-1}(n_{a}\cdot n_{b})$. The following definition, due
to KS, satisfies our requirements and serves as a basis for handling the
two-qubit case. It is based on the fact that each unit vector divides $S^{2}$
in two hemispheres. We define $\rho _{a}(\lambda )$ as being different from
zero only on the intersection of the northern hemispheres of $n_{a}$ and $%
n_{\lambda }=\lambda $, where it takes the value $n_{\lambda }\cdot
n_{a}/\pi $. This can be expressed with the help of Heaviside's
step-function ($\Theta (x)=1$, for $x\geq 0$ and $\Theta (x)=0$, for $x<0$)
as

\begin{equation}
\rho _{a}(\lambda )=\frac{n_{\lambda }\cdot n_{a}}{\pi }\Theta \left(
n_{\lambda }\cdot n_{a}\right) \text{.}  \label{2}
\end{equation}
On the other hand, the characteristic functions are defined as

\begin{equation}
\chi _{i}^{\pm }(\lambda )=\Theta \left( n_{\lambda }\cdot n_{i}^{\pm
}\right) ,\text{ \ \ }i=a,b.  \label{3}
\end{equation}
Here, $n_{b}^{\pm }$ - similarly to $n_{a}^{\pm }$ - are in one-to-one
correspondence with the eigenvectors $\left| \psi _{b}^{\pm }\right\rangle $
of Bob's observable $\widehat{B}=b_{0}+\mathbf{b}\cdot \mathbf{\sigma }$.
Next, we show that $\int \rho _{a}(\lambda )\chi _{b}^{+}(\lambda )d\lambda
=\cos ^{2}\left( \theta _{ab}/2\right) $ and $\int \rho _{a}(\lambda )\chi
_{b}^{-}(\lambda )d\lambda =\sin ^{2}\left( \theta _{ab}/2\right) $, as
desired. The measure is taken to be $d\lambda \equiv dS_{\lambda }=\sin
\theta _{\lambda }d\theta _{\lambda }d\varphi _{\lambda }$, the surface
element on the unit sphere.

Let us take $n_{a}=n_{a}^{+}\leftrightarrow \left| \psi
_{a}^{+}\right\rangle $ and $n_{b}=n_{b}^{+}\leftrightarrow \left| \psi
_{b}^{+}\right\rangle $, for concreteness. Notice that, because $%
\left\langle \psi _{a}\right| \widehat{\Pi }_{b}\left| \psi
_{a}\right\rangle =\left\langle \psi _{b}\right| \widehat{\Pi }_{a}\left|
\psi _{b}\right\rangle $, it should hold true that $I_{ab}\equiv \int \rho
_{a}(\lambda )\chi _{b}(\lambda )d\lambda =\int \rho _{b}(\lambda )\chi
_{a}(\lambda )d\lambda \equiv I_{ba}$. Writing $I_{ab}$ more explicitly we
obtain

\begin{equation}
I_{ab}=\frac{1}{\pi }\int n_{\lambda }\cdot n_{a}\Theta \left( n_{\lambda
}\cdot n_{a}\right) \Theta \left( n_{\lambda }\cdot n_{b}\right) d\lambda =%
\frac{1}{\pi }\int_{N_{a}\cap N_{b}}n_{a}\cdot n_{\lambda }dS_{\lambda },
\label{4}
\end{equation}
where $N_{a}\cap N_{b}\equiv S_{ab}$ is the area over which the integration
is effectively restricted by $\Theta \left( n_{\lambda }\cdot n_{a}\right)
\Theta \left( n_{\lambda }\cdot n_{b}\right) $. $N_{a}$ and $N_{b}$ are the
northern hemispheres belonging to the Poles $n_{a}$ and $n_{b}$,
respectively. Now, the last expression in Eq.(\ref{4}) expresses $I_{ab}$ as
a flux integral. Defining the vector-field $v_{a}(r)=n_{a}\times r/2$, we
have $n_{a}=\nabla \times v_{a}(r)$, so that, by applying Stoke's theorem,
we obtain

\begin{eqnarray}
I_{ab} &=&\frac{1}{\pi }\int_{S_{ab}}\nabla \times v_{a}(r)\cdot n_{\lambda
}dS_{\lambda }=\frac{1}{\pi }\oint_{\partial S_{ab}}v_{a}(r_{\lambda })\cdot
dr_{\lambda }  \label{5} \\
&=&\frac{1}{2\pi }\oint_{\partial S_{ab}}n_{a}\times r_{\lambda }\cdot
dr_{\lambda }=\frac{1}{2\pi }\oint_{\partial S_{ab}}\left( r_{\lambda
}\times \frac{dr_{\lambda }}{ds}\cdot n_{a}\right) ds,  \notag
\end{eqnarray}
where $\partial S_{ab}$ means the contour limiting $S_{ab}$ and $ds$ is the
arc-length used to parameterize the curve $r_{\lambda }(s)\equiv n_{\lambda
}(s)$ on $S^{2}$, so that $dr_{\lambda }/ds$ is a unit-vector tangent to the
sphere. The contour $\partial S_{ab}$ limiting $S_{ab}$ is made of two great
circles, $C_{a}$ and $C_{b}$, the ``equators'' relative to $n_{a}$ and $%
n_{b} $, respectively. They intersect at two antipodal points, $P_{1}$ and $%
P_{2}$, say. The contour integral can thus be split in two line integrals,
one going from $P_{1}$ to $P_{2}$ along $C_{a}$, and the other from $P_{2}$
back to $P_{1}$ along $C_{b}$. Each of these curves is half a great-circle
and has thus a length equal to $\pi $. Now, $r_{\lambda }\times dr_{\lambda
}/ds$ is also a unit vector - the so-called bivector in the theory of curves
- which equals $n_{a}$ along $C_{a}$ and $n_{b}$ along $C_{b}$. Whence,

\begin{eqnarray}
I_{ab} &=&\frac{1}{2\pi }\left( \int_{C_{a}}n_{a}\cdot
n_{a}ds+\int_{C_{b}}n_{b}\cdot n_{a}ds\right) =  \label{6} \\
&=&\frac{1}{2}\left( 1+n_{b}\cdot n_{a}\right) =\cos ^{2}\left( \frac{\theta
_{ab}}{2}\right) .  \notag
\end{eqnarray}

The case $n_b=n_b^{-}=-n_b^{+}$ gives $I_{ab}=\left( 1-n_b\cdot n_a\right)
/2=\sin ^2\left( \theta _{ab}/2\right) $, and this completes the proof. Our
procedure also makes clear how the symmetry under $n_a\leftrightarrow n_b$
arises, so that $I_{ab}=I_{ba}$, as already mentioned.

\subsection{A Kochen-Specker model for the two-qubit case\label{subsection2}}

Let us now turn to the two-qubit case, specifically addressing the
experiment reported by Gr\"{o}blacher \emph{et al} \cite{gröblacher}. Our
aim is to give a counterexample for a ``no-go'' assertion which states that
no contextual HV model - within a wide class - would be capable of
explaining the results of the said experiment. If this assertion were true,
then a wide class of contextual HV models would have been ruled out by the
experiment. As already mentioned, this experiment was conceived as a test of
a Leggett-type inequality that was derived from the following assumptions:
(i) In experiments using a source that emits pairs of photons with
well-defined polarizations $n_{u}$ and $n_{v}$, each emitted pair belongs to
a subensemble that is defined through a density $\rho _{uv}$. Measurements
performed by Alice and Bob - possibly influencing each other in a non-local
way - produce outputs obeying Malus' law, Eq.(\ref{1}). (ii) For a general
source producing mixtures of polarized photons, there is a function $F(u,v)$%
, ruling the distribution of polarizations. Under these circumstances, a
Leggett-type inequality should hold true \cite{gröblacher}. However, QM
violates this inequality and is in accordance with the results obtained in
the experiment \cite{gröblacher}. The source that was used in the experiment
produced polarization-entangled singlet states $\left| \Psi
^{-}\right\rangle _{AB}=\left( \left| H\right\rangle _{A}\left|
V\right\rangle _{B}-\left| V\right\rangle _{A}\left| H\right\rangle
_{AB}\right) /\sqrt{2}$ of vertically ($V$) and horizontally ($H$) polarized
photons, by means of a standard type-II parametric down-conversion process.
We construct next a contextual model satisfying the above requirements and
being in accordance with the predictions of QM. This model will be tailored
so as to reproduce the results of the particular experiment we have in
sight, i.e., we will assume an initial distribution that corresponds to the
singlet state $\left| \Psi ^{-}\right\rangle _{AB}$. To this end, let us
first set $\rho _{uv}(\lambda _{1},\lambda _{2})=\rho _{u}^{+}(\lambda
_{1})\rho _{v}^{+}(\lambda _{2})$, where $\rho _{u}^{+}$, $\rho _{v}^{+}$
are defined as in Eq.(\ref{2}). We will see that this choice enforces Malus'
law. Thereafter, we will choose an appropriate contextual distribution $%
F(u,v)$ that describes the initial state. Note that we have divided the HVs
in two groups: $\lambda =(\lambda _{1},\lambda _{2})$. A possible
interpretation of this division is that $\lambda _{1}$ relates to Alice's
measuring device, and $\lambda _{2}$ to Bob's. This interpretation would be
consistent with the assumption of contextual influences acting upon the
source. Alternatively, one could take $\lambda _{1,2}$ to be parameters that
are carried by the particles that are registered by Alice and Bob. In such a
case, $\rho _{uv}$ would have been locally defined. It is straightforward to
see that our $\rho _{uv}$ satisfies Malus' law. Indeed, $\int d\lambda
_{2}d\lambda _{1}\rho _{uv}\chi _{a}^{\pm }(\lambda _{1})=\int d\lambda
_{2}\rho _{v}^{+}(\lambda _{2})\int d\lambda _{1}\rho _{u}^{+}(\lambda
_{1})\chi _{a}^{\pm }(\lambda _{1})=(1\pm n_{u}.n_{a})/2$, so that $%
\overline{A}=\int d\lambda \rho _{uv}A(\lambda _{1})=\int d\lambda \rho
_{uv}(\chi _{a}^{+}(\lambda _{1})-\chi _{a}^{-}(\lambda _{1}))=n_{u}.n_{a}$.
Here, we have used the results of the first part, i.e., that $\rho
_{u}^{+}(\lambda )$ is normalized, and Eq.(\ref{6}) with $n_{b}\rightarrow
n_{u}$. Similarly, one obtains $\overline{B}=\int d\lambda \rho
_{uv}B(\lambda _{2})=n_{v}.n_{b} $. Because our $\rho _{uv}$ factorizes,
then $\overline{AB}(u,v)=\overline{A}\cdot \overline{B}%
=(n_{u}.n_{a})(n_{v}.n_{b})$, so that

\begin{equation}
E_{ab}=\left\langle AB\right\rangle =\int dudvF(u,v)\overline{AB}(u,v)=\int
dudvF(u,v)(n_{u}.n_{a})(n_{v}.n_{b}).  \label{7}
\end{equation}
Let us now take a contextual distribution function which is appropriate for
our scopes. A possible choice is the following one:

\begin{equation}
F_{ab}(u,v)=\frac{1}{2\pi ^{2}}\left( \chi _{a}^{+}(\lambda _{u})\chi
_{a}^{-}(\lambda _{v})+\chi _{b}^{-}(\lambda _{u})\chi _{b}^{+}(\lambda
_{v})\right) .  \label{8}
\end{equation}
With this $F_{ab}(u,v)$ we obtain the desired result, i.e., that $%
E_{ab}=\left\langle AB\right\rangle =-n_{a}.n_{b}$, in accordance with the
quantum-mechanical prediction. Indeed, replacing Eq.(\ref{8}) in Eq.(\ref{7}%
) $E_{ab}$ can be written as the sum of two terms: $E_{ab}=(I_{a}+I_{b})/2$,
with $I_{j}=\int d\lambda _{u}d\lambda _{v}(n_{u}.n_{a})(n_{v}.n_{b})\chi
_{j}^{+}(\lambda _{u})\chi _{j}^{-}(\lambda _{v})/\pi ^{2}$, $(j=a,b)$. We
can calculate the two integrals following a similar procedure as we did
before:
\begin{eqnarray}
I_{a} &=&\frac{1}{\pi ^{2}}\int d\lambda _{u}d\lambda
_{v}(n_{u}.n_{a})(n_{v}.n_{b})\chi _{a}^{+}(\lambda _{u})\chi
_{a}^{-}(\lambda _{v})=\overset{1}{\overbrace{\frac{1}{\pi }\int d\lambda
_{u}\chi _{a}^{+}(\lambda _{u})(n_{u}.n_{a})}}\int d\lambda _{v}(n_{v}.n_{b})%
\frac{\chi _{a}^{-}(\lambda _{v})}{\pi }  \notag \\
&=&\frac{1}{\pi }\int_{S_{a}}\nabla \times v_{b}(r_{\lambda })\cdot
n_{\lambda }dS_{\lambda }=\frac{1}{2\pi }\oint_{C_{a}}n_{b}\times r_{\lambda
}\cdot dr_{\lambda }=\frac{1}{2\pi }\oint_{C_{a}}r_{\lambda }\times \frac{%
dr_{\lambda }}{ds}\cdot n_{b}ds=-n_{a}.n_{b}.  \label{9}
\end{eqnarray}
Here, $S_{a}$ is the southern hemisphere of $n_{a}$, $C_{a}$ its equator,
and $v_{b}(r)=n_{b}\times r/2$. We have applied Stokes' theorem as in Eq.(%
\ref{5}), but now the contour $C_{a}$ is oriented clockwise. In a similar
way one obtains $I_{b}=\int d\lambda _{u}d\lambda
_{v}(n_{u}.n_{a})(n_{v}.n_{b})\chi _{b}^{+}(\lambda _{u})\chi
_{b}^{-}(\lambda _{v})/\pi ^{2}=-n_{a}.n_{b}$, so that $E_{ab}=\left\langle
AB\right\rangle =(I_{a}+I_{b})/2=-n_{a}.n_{b}$, as desired. Note that by
taking $F_{a}(u,v)=\left( \chi _{a}^{+}(\lambda _{u})\chi _{a}^{-}(\lambda
_{v})\right) /\pi ^{2}$ we would have obtained the same result. We chose $%
F_{ab}$ for the sake of symmetry between the two parties. In any case, we
have succeeded in constructing a counterexample, i.e., a contextual HV model
that reproduces the predictions of QM for a particular case, viz, the
experiment performed by Gr\"{o}blacher \emph{et al}.

Although we have worked out in detail a particular case, it is clear that
our procedure could be extended so as to mimic QM in a general case. That
is, for cases in which the initial state is not the singlet-state, we could
also construct a contextual HV\ model. The singlet-state is one of the four
states that constitute the standard Bell-basis, which is given by $\left|
\Psi ^{\pm }\right\rangle =\left( \left| H\right\rangle _{A}\left|
V\right\rangle _{B}\pm \left| V\right\rangle _{A}\left| H\right\rangle
_{AB}\right) /\sqrt{2}$ and $\left| \Phi ^{\pm }\right\rangle =\left( \left|
V\right\rangle _{A}\left| V\right\rangle _{B}\pm \left| H\right\rangle
_{A}\left| H\right\rangle _{AB}\right) /\sqrt{2}$ . We have seen that $%
\left\langle \Psi ^{-}\right| AB\left| \Psi ^{-}\right\rangle =-n_{a}.n_{b}$%
. For the other Bell states, $\left\langle AB\right\rangle $ can be
expressed in terms of the components of $n_{a}=(a_{x},a_{y},a_{z})$ and $%
n_{b}=(b_{x},b_{y},b_{z})$ as $\left\langle \Psi ^{+}\right| AB\left| \Psi
^{+}\right\rangle =a_{x}b_{x}+a_{y}b_{y}-a_{z}b_{z}$, $\left\langle \Phi
^{+}\right| AB\left| \Phi ^{+}\right\rangle
=a_{x}b_{x}-a_{y}b_{y}+a_{z}b_{z} $, and $\left\langle \Phi ^{-}\right|
AB\left| \Phi ^{-}\right\rangle =-a_{x}b_{x}+a_{y}b_{y}+a_{z}b_{z}$. Now, it
is clear from our above results that we can easily choose an appropriate
distribution $F_{ab}(u,v)$ for all Bell states, as we did for $\left| \Psi
^{- }\right\rangle$. Indeed, take for example the state $\left| \Psi
^{+}\right\rangle $. We can write $\left\langle \Psi ^{+}\right| AB\left|
\Psi ^{+}\right\rangle =\left( n_{a}.e_{x}\right) \left( n_{b}.e_{x}\right)
+\left( n_{a}.e_{y}\right) \left( n_{b}.e_{y}\right) -\left(
n_{a}.e_{z}\right) \left( n_{b}.e_{z}\right) $, with $e_{x}$, $e_{y}$, $%
e_{z} $ being the unit vectors with respect to which we have defined the
coordinates of $n_{a}$ and $n_{b}$. Having expressed $\left\langle \Psi
^{+}\right| AB\left| \Psi ^{+}\right\rangle $ in terms of scalar products,
it is straightforward to choose $F_{ab}(u,v)$ by looking at the derivation
of Eq.(\ref{9}). Indeed, from an analogous calculation we can readily prove
that $\int d\lambda _{u}d\lambda _{v}(n_{u}.n_{a})(n_{v}.n_{b})\chi
_{i}^{+}(\lambda _{u})\chi _{j}^{\pm }(\lambda _{v})/\pi ^{2}=\pm \left(
n_{a}.e_{i}\right) (n_{b}.e_{j})$, $i,j=x,y,z$. Hence, we can choose $%
F_{ab}(u,v)$ so as to obtain any desired combination of scalar products when
we insert it into Eq.(\ref{7}). A general, initial state $\left| \Psi
\right\rangle $ can be written as a linear combination of the Bell states.
One can then easily check that $\left\langle \Psi \right| AB\left| \Psi
\right\rangle $ contains only binary products of the Cartesian components of
$n_{a}$ and $n_{b}$. Thus, the above result applies for the general case of
an arbitrary initial state $\left| \Psi \right\rangle $. The case of an
initial mixed state can be dealt with by writing a combination of different
distributions $F_{ab}(u,v)$ with appropriate weights.

\section{Conclusions\label{section2}}

In view of the two-qubit model we have discussed, we can draw the following
conclusions. Our model reproduces the predictions of QM for the experiment
of Gr\"{o}blacher \emph{et al}. and hence violates the Legget-type
inequality derived in Ref.\cite{gröblacher}. Though one can qualitatively
consider other, simpler realistic non-local models that are not addressed by
the Leggett inequality \cite{aspect}, the one presented here is an explicit
one, that is very akin to the ones considered by Leggett and by
Gr\"{o}blacher \emph{et al}. The point of departure from the derivation
presented in Ref.\cite{gröblacher} is that we considered a contextual
distribution $F_{ab}(u,v)$ instead of the non-contextual $F(u,v)$ that was
assumed in Refs. \cite{gröblacher,leggett}. Within a contextual theory it is
certainly justified to take $F_{ab}(u,v)$ together with contextual densities
$\rho _{uv}(\lambda ,a,b)$. Moreover, as already mentioned, contextual
densities or distributions would not be extraneous to a classical approach.
They could be thought of as arising from solving some fundamental partial
differential equations. To prescribe ``boundary conditions'' in order to
solve these equations would be tantamount to allow ``contextuality''. We can
therefore conclude that the experiments recently reported by Gr\"{o}blacher
\emph{et al }\cite{gröblacher} do not address a broad plausible class of
contextual, hidden-variable models. They did address a class whose defining
feature requires that models pertaining to it may ascribe contextual
qualities to the measuring devices, but not to those devices that were
included as part of a ``source''. Being deliberately provocative, let us
illustrate the said defining feature by referring to quantum optics
experiments. Advocates of the aforementioned class must contend that even
though the action of a calcite crystal (in front of a detector) could be
affected by distant devices, the action of a beta-barium-borate crystal (in
a source) could not. This can hardly be an assumption worth to be tested.
Instead of conducting experiments of increased refinement \cite{gröblacher2}
to test realistic models of a class that could have been discarded from the
outset, it would be more meaningful to introduce, for instance, variable
polarizers as a tool for testing non-locality. Consider for instance an
experiment with variable polarizers that yields the results predicted by QM.
The model presented here could be slightly modified so as to explain such
results, but at the cost of entering into the class of highly
counterintuitive ones. Let us finally mention that by properly testing
contextuality in physically plausible HV models one could complement some
recently proposed experiments that address determinism together with
non-contextuality \cite{fdz,malley}. If these experiments do confirm the
predictions of QM, then contextuality should be necessarily included in any
fundamental description of physical phenomena, unless we are ready to admit
an inherent indeterminacy in the ultimate nature of these phenomena. In
other words, we would be faced with the choice between contextuality and
indeterminism.

\end{document}